\documentstyle[prb,aps]{revtex}
\begin{document}
\title{Charge density waves and surface Mott insulators 
for adlayer structures on semiconductors: extended Hubbard modeling} 
\vspace{10mm}
\author{Giuseppe Santoro$^{1,2}$, Sandro Scandolo$^{1,2}$, and 
Erio Tosatti$^{1,2,3}$}
\address{
$^{(1)}$ International School for Advanced Studies (SISSA), Via Beirut 2,
Trieste, Italy\\
$^{(2)}$ Istituto Nazionale per la Fisica della Materia (INFM), 
Via Beirut 2, Trieste, Italy\\
$^{(3)}$ International Center for Theoretical Physics (ICTP), Strada Costiera,
Trieste, Italy}
\maketitle
\vspace{10mm}
\begin{abstract}
Motivated by the recent experimental evidence of commensurate surface
charge-density-waves (CDW) 
in Pb/Ge(111) and Sn/Ge(111) $\surd{3}$-adlayer structures, 
as well as by the insulating states found on K/Si(111):B and SiC(0001),
we have investigated the role of electron-electron interactions, and also of 
electron-phonon coupling, on the narrow surface state band originating
from the outer dangling bond orbitals of the surface.
We model the  $\surd{3}$ dangling bond lattice by an extended
two-dimensional Hubbard model at half-filling on a triangular lattice. 
The hopping integrals are calculated by fitting first-principle results for
the surface band.
We include an on-site Hubbard repulsion $U$ and a nearest-neighbor Coulomb
interaction $V$, plus a long-ranged Coulomb tail.
The electron-phonon interaction is treated in the deformation potential
approximation. We have explored the phase diagram of this 
model including the possibility of commensurate $3\times 3$ phases, 
using mainly the Hartree-Fock approximation. 
For $U$ larger than the bandwidth we find a non-collinear antiferromagnetic
SDW insulator, possibly corresponding to the situation on the SiC 
and  K/Si surfaces. For $U$ comparable or smaller, a rich phase 
diagram arises, with several phases involving
combinations of charge and spin-density-waves (SDW), with or without a net
magnetization. We find that insulating, or partly metallic $3\times 3$ CDW 
phases can be stabilized by two different physical mechanisms. One is the
inter-site repulsion $V$, that together with electron-phonon coupling can 
lower the energy of a charge modulation.
The other is a novel magnetically-induced Fermi surface nesting,
stabilizing a net cell magnetization of 1/3, plus a collinear SDW, plus
an associated weak CDW. Comparison with available experimental 
evidence, and also with first-principle calculations is made.
\end{abstract}
 

\section{Introduction}

Pb and Sn $\sqrt{3}\times \sqrt{3}$ adlayer structures on the (111)
surface of Ge have recently revealed a reversible charge density wave 
(CDW) transition to a low temperature reconstructed $3\times 3$ phase.
\cite{nature,Modesti_1,Modesti_2,Avila,LeLay,Sn_carpinelli,asensio} 
A half-filled surface state band makes the high temperature phase metallic.
The low temperature phase is either metallic -- as seems to be the case
for Sn/Ge(111) -- or weakly gapped, or pseudo-gapped, as suggested
for Pb/Ge(111). 

Related isoelectronic systems, like the $\sqrt{3}$-adlayer of Si on
the (0001) surface of SiC \cite{SiC} and on K/Si(111):B,\cite{KSi}
show a clear insulating behavior, with a large gap,
no structural anomalies, no CDWs, and no transitions, at least to
our present knowledge.

These adsorbate surfaces are altogether mysterious. 
The very existence of a $\sqrt{3}\times \sqrt{3}$ adsorbate phase, 
with coverage 1/3, is puzzling. 
For isoelectronic Si on Si(111), or Ge on Ge(111), for instance, there exists 
no such phase. The stable low-coverage phase are $7\times 7$ and $c(2\times 8)$ 
respectively, whose coverage is instead close to 1/4. 
They are made up of $2 \times2$ basic building blocks, each with one adatom
saturating three out of four first-layer atoms, and one unsaturated
first-layer atom, the ``restatom''. In this adatom-restatom block, 
the nominally unpaired electron of the adatom and that of the restatom pair 
off together, giving rise to a stable, fully saturated, insulating surface. 
By contrast, the $\sqrt{3}\times \sqrt{3}$ phases
with 1/3 coverage are very common for trivalent adsorbates, such as
Ga and In, and for pentavalent ones like As, on the same (111) surfaces.
These adatoms lack the unpaired electron, and can therefore lead to
a fully saturated insulating surface without the need for any restatoms. 

A $\sqrt{3}\times \sqrt{3}$ adsorbate phase of {\it tetravalent} adatoms  
is bound by construction to possess one unpaired electron per adatom,
giving rise to a very destabilizing half-filled metallic surface state band.
Seen in this crude light, it is a puzzle why this kind of coverage should  
constitute even only a locally stable state of the surface.

Looking more closely, we may speculate that SiC(0001)\cite{SiC} 
and K/Si(111):B,\cite{KSi} most likely Mott-Hubbard insulators,
\cite{santoro,Northrup,Anisimov,KSi} are perhaps ``stabilized'' by Coulomb 
repulsions, so large to make it anyway difficult for electrons to move. 
For the more innocent-looking, less correlated, Pb/Ge(111) and Sn/Ge(111), 
this argument is less obvious, and the puzzle remains. The function of 
the $3\times 3$ CDW state -- whatever its real nature -- most likely 
serves the function of stabilizing these otherwise unstable surfaces at 
low temperatures. Nonetheless, the CDW periodicity chosen by the surface CDW
-- $3\times 3$, meaning a $\sqrt{3}\times \sqrt{3}$ super-cell of adatoms -- 
is not at all evident. In fact, it replaces a supposedly unstable state 
characterized by an odd number of electrons/cell (three), with another 
where the electron number (nine) is, alas, odd again.

Be all that as it may, there is little doubt that the main factor driving the
phenomena on all these surfaces, appears to be precisely the half-filled--
and extremely narrow -- surface state band.  We thus
begin with a discussion that in principle encompasses all the
$\sqrt{3}\times \sqrt{3}$ tetravalent adsorbed surfaces.

We believe the following points to be of general validity:

{\em i)\/} {\bf Poor nesting}. 

Two-dimensional Fermi surface (FS) nesting in the half-filled surface 
states\cite{tosatti} has been repeatedly invoked as the driving 
mechanism for the
CDW instability in the case of Pb/Ge,\cite{nature,asensio} but excluded 
for the case of Sn/Ge.\cite{Sn_carpinelli,Sn_scandolo} 
However, by inspecting either photoemission $k(E)$ data,
\cite{Modesti_2,Avila,LeLay,asensio} and existing 
first-principle (LDA) calculations\cite{nature,Sandro,Sn_carpinelli}
of the surface half-filled band (the``adatom dangling 
bond band''), we fail to detect a particularly good nesting of the 
two-dimensional FS at the surface Brillouin zone (BZ) corner 
${\bf K}=(4\pi/3a,0)$. The wavevector-dependent susceptibility 
generated by the calculated band structure, in particular, has no 
especially large value at this k-point, and rather peaks elsewhere 
(see inset in Fig.\ \ref{band_bz_chi0:fig}). 
To be sure, there is nothing preventing in general a good nesting at 
${\bf K}=(4\pi/3a,0)$, or any other k-point.
However, insofar as the surface state band is really lying in a bulk gap at 
each single k-point, it should be with good accuracy 
-- by simple state counting and charge neutrality --
precisely half filled. 
This implies that the filled and empty state areas should be equal. 
Hypothetical Fermi surfaces with this kind of shape and good nesting at 
${\bf K}=(4\pi/3a,0)$ do not appear to be compatible 
with an integer electron number.
We thus believe lack of perfect nesting to be the case for both 
Pb/Ge as for Sn/Ge.

Fig.\ \ref{band_bz_chi0:fig}, showing a tight binding fit to the LDA surface 
band dispersion for the test-case of Si(111)/Si,\cite{Sandro} as well as the
corresponding FS and Lindhard density response function $\chi_o(\bf q)$,
\[ \chi_o(\bf q) \,=\, \int_{BZ} \frac{d\bf k}{(2\pi)^2}
\; \frac{ n_{\bf k} - n_{{\bf k}+{\bf q}} }
{\epsilon_{{\bf k}+{\bf q}} - \epsilon_{\bf k} }  \;, \]
$n_{\bf k}$ and $\epsilon_{\bf k}$ being the occupation number and energy 
of an electron with Bloch momentum ${\bf k}$, 
provides a concrete illustration of these statements.
We note, in passing, that a strong nesting at ${\bf K}$ is, on the contrary, 
automatically guaranteed if the surface band acquires a uniform magnetization
in such a way that the densities of up and down electrons become, 
respectively, $2/3$ and $1/3$.\cite{Sandro} 
The majority spins would then fill the region external to the reduced BZ in 
Fig.\ \ref{band_bz_chi0:fig}, and their FS would be strongly nested. 
This suggestion, which turns out to be correct at the mean-field level,  
points into the direction of a possible role played by magnetism in these 
systems.  

{\em ii)\/} {\bf Importance of electron-electron interactions}. 

The width $W$ of the surface band is relatively small:
$W\approx 0.5$ eV for Pb and Sn/Ge(111), $W\approx 0.3$ eV for SiC(0001). 
Moreover, this band is half-filled.
These facts call for a careful consideration of electron-electron
interactions, as well as of electron-phonon (e-ph), as possible sources of 
instability. The importance of electron-electron interaction is 
underlined by the different phenomenology
of SiC(0001) and K/Si(111):B with respect to Pb-Sn/Ge(111). 
The stronger insulating character of the former surfaces parallels
closely their stronger electron-electron 
repulsions, connected both with more localized surface Wannier functions
(see later on),
and with reduced screening, due to larger bulk semiconducting gaps.

{\em iii)\/} {\bf Weakness of LDA calculations for ground state prediction}. 

LDA electronic structure calculations -- an extremely well tested tool
in many areas-- are certainly suitable for a weakly interacting system,
such as the bulk semiconductor, or a passivated semiconductor surface. 
They are less reliable, especially when they do not include spin, in 
predicting the stable state and the instabilities of a narrow band system. 
For instance, the phenomenology of SiC(0001) -- suggesting a Mott-Hubbard 
insulator -- is unreproducible by LDA. 
The onset of a CDW on Sn/Ge(111)is also not predicted by recent 
LDA calculations.\cite{Sn_carpinelli,Sn_scandolo} While there is no
reason to doubt the basic credibility of the one-electron band energies 
obtained from these Kohn-Sham equations, the mean-field treatment of 
interactions, the screened local exchange, and especially 
the neglect of magnetic
correlations are the standard source of problems with LDA.  As a
consequence, it will be necessary to worry more substantially about
interactions, and to use methods which, even if mean-field, permit the
inclusion of strong correlations, including magnetic effects.

{\em iv)\/} 
{\bf Interaction-driven mechanisms for $3\times 3$ CDW instabilities}.

There are several different couplings which the surface electrons, as they hop
weakly between a surface adatom site and another, experience and can influence the 
formation of the CDW, or of an insulating ground state:
a) on-site, and nearest-neighbor (n.n.) inter-site electron-electron repulsion; 
b) on-site effective attraction (negative Hubbard-$U$ term) of electron-phonon
origin.

Because of poor nesting, electron-phonon alone is unlikely to 
drive the $3\times 3$ CDW. 
At weak coupling, the susceptibility peak in Fig.\ \ref{band_bz_chi0:fig} would
rather drive an incommensurate periodicity. 
At strong coupling, the frustration associated to the triangular lattice, 
will favor, in general, a superconducting ground state over a 
CDW phase (see Appendix).\cite{santos}

On the other hand, the electron-electron interaction, both on-site and, 
independently, nearest neighbor, naturally suggests, as we shall see later, 
the $3\times 3$ surface periodicity, which is found experimentally. 

The approach we will take is based on an extended Hubbard-Holstein model.
It is by necessity a ``non-first-principle'' approach and, as such,
has no strong predictive power. 
However, it is made more realistic by using parameters extracted 
from first-principle calculations, and we find it very helpful in clarifying 
the possible scenarios as a function of the strength of electron-electron 
interactions. 
Because of this rather qualitative use, we will make no attempt to push the 
accuracy of treatment of this model to a very high level of sophistication. 
The basic tool will be the unrestricted Hartree-Fock approximation. 
Although mean field, it allows magnetic solutions, favored by exchange which 
is unscreened. 

\section{Model}
Each tetravalent adatom on a (111) semiconductor surface carries a dangling
bond -- an unpaired electron in an unsaturated orbital.
In the $\sqrt{3}\times\sqrt{3}$ structure, the dangling bonds of the adatoms 
give rise to a band of surface states which lies in the bulk semiconductor
gap. \cite{nature,Sandro} By electron counting, such a band is half-filled. 
Our basic starting point is the quantitatively accurate surface state band 
dispersion $\epsilon_{\bf k}$ which one calculates in gradient-corrected LDA,
\cite{nature,Sandro}. 
It is shown in Fig.\ \ref{band_bz_chi0:fig} for the case of Si/Si(111). 
The solid and dashed lines in Fig.\ \ref{band_bz_chi0:fig} are tight-binding 
fits to the LDA results obtained by including, respectively, up to the $6^{th}$ 
and up to the $2^{nd}$ shell of neighbors.
The fit with hopping integrals $t_1,t_2,\cdots,t_6$ is quite good.
Less good, but qualitatively acceptable, is the fit obtained using only
nearest neighbor (n.n.) and next-nearest neighbor (n.n.n.) hopping integrals
$t_1$ and $t_2$.
The Fermi surface (FS) for the half-filled surface band is shown in the
upper inset of Fig.\ \ref{band_bz_chi0:fig}.
It is important to stress that the FS does not show good nesting properties
at the wavevector ${\bf q}={\bf K}$ (the BZ corner).
This feature is shared by all LDA calculations on similar 
systems.\cite{nature,Sandro,Sn_carpinelli}
Albeit small, the bandwidth $W$ of the surface band is much greater than one 
would predict by a direct overlap of adatom dangling bonds, as the adatoms
are very widely apart, for instance about $7\AA$ on Ge(111). Hopping is
indirect, and takes place from the adatom to the first-layer atoms
underneath, from that to a second-layer atom, then again to a first-layer
atom underneath the other adatom, and from there finally to other adatom
dangling bond. 
Thus, when expressed in terms of elementary hopping processes between 
hybrid orbitals, electron hopping between two neighboring adatom dangling 
bonds is fifth order.
As a result, the final dispersion of the surface state band strongly 
parallels that of the closest bulk band, the valence band. Correspondingly,
hybridization effects of the dangling bond orbitals with first, second, and
even third, bulk layer orbitals are strong, as shown by the extension 
into the bulk of the Wannier orbital associated to the LDA surface band 
(Fig.\ \ref{wannier:fig}).

In spite of this, we can still associate to every adatom a Wannier orbital 
and write the effective Hamiltonian for the surface band as follows:
\begin{equation}
H \,=\, \sum_{{\bf k}}^{BZ} \sum_{\sigma} \epsilon_{\bf k} 
c^{\dagger}_{{\bf k},\sigma} c_{{\bf k},\sigma} \,+\, H_{\rm ph} \,+\, 
H_{\rm e-ph} \,+\, H_{\rm int} \;,
\end{equation}
where $c^{\dagger}_{{\bf k},\sigma}$ is the Fourier transform of 
the Wannier orbital, namely the surface state in a Bloch picture.
The sum over the wavevectors runs over the surface BZ.
$H_{\rm int}$ includes correlation effects which are not correctly accounted 
for within LDA, which we parametrize as follows:
\begin{equation}
H_{\rm int} = U \sum_{{\bf r}} n_{{\bf r},\uparrow} n_{{\bf r},\downarrow}
+ \frac{1}{2} \sum_{{\bf r}\ne {\bf r}'} V_{{\bf r}-{\bf r}'} 
(n_{{\bf r}}-1) (n_{{\bf r}'}-1) \;.
\end{equation}
Here $U$ is an effective repulsion (Hubbard-$U$) for two electrons on the same 
adatom Wannier orbital, and $V_{{\bf r}-{\bf r}'}$ is the 
direct Coulomb interaction between different sites ${\bf r}$ and 
${\bf r}'$.\cite{non-diag:nota}
Let $V$ be the n.n.\ value of $V_{{\bf r}-{\bf r}'}$, which is, clearly, the
largest term.
We have considered two models for $V_{{\bf r}-{\bf r}'}$: 
a model (A) in which we truncate
$V_{{\bf r}-{\bf r}'}$ to n.n., and a model (B) in which
$V_{{\bf r}-{\bf r}'}$ has a long range Coulombic tail of the form
\[ V_{{\bf r}-{\bf r}'} \,=\, \frac{a V}{|{\bf r}-{\bf r}'|} \;, \]
where $a$ is the n.n.\ distance.
The results for model B are qualitatively similar to those of A, 
and will be only briefly discussed later on. In other words, even if
most of the detailed results in this paper will be base on the n.n.\ 
$V_{{\bf r}-{\bf r}'}$, their validity is more general. 

LDA estimates of the {\it bare} coulomb repulsion $U_o$ and $V_o$ 
between two electrons respectively on the same and on neighboring Wannier 
orbitals are -- for our test case of Si(111)/Si -- of about $3.6$ eV and 
$1.8$ eV respectively.\cite{Sandro} 
Screening effects by the the underlying bulk are expected to
reduce very substantially these repulsive energies. An order of magnitude
estimate for $U$ and $V$ is obtained by dividing their bare values
by the image-charge screening factor, $(\epsilon +1)/2\approx 6$, 
yielding, for Si, $U=0.6$ eV ($10 t_1$), and $V=0.3$ eV ($5 t_1$).
Corresponding values would be somewhat smaller for Ge(111), in view
of a very similar dispersion \cite{Sn_scandolo} and of a ratio of about 
4/3 between the dielectric constants of Ge and Si. SiC(0001), the opposite
is true. The surface state band is extremely narrow, of order $0.3$ eV
\cite{pollmann}, while the bulk dielectric constant is only about $6.5$. 

As for the e-ph interaction, in principle both the on-site
Wannier state energy and the hopping matrix elements between neighbors
depend on the positions of the adatoms. 
Within the deformation potential approximation, we consider only a 
linear dependence of the on-site energy from a single ionic coordinate 
(for instance, the height $z_{\bf r}$ of the adatom measured 
from the equilibrium position), and take 
\begin{equation} \label{e-ph-ham:eqn}
H_{\rm e-ph} = -g \sum_{{\bf r}} z_{\bf r} (n_{\bf r}-1) \;,
\end{equation}
with $g$ of the order of $\approx 1$ eV/$\AA$. The free-phonon term will
have the usual form
\begin{equation}
H_{\rm ph} = \sum_{\bf k}^{BZ} \hbar \omega_{\bf k}
\left( b^{\dagger}_{\bf k} b_{\bf k} + \frac{1}{2} \right) \;,
\end{equation}
%
%
where $b_{\bf k}$ is the phonon annihilation operator, and 
$\hbar \omega_{\bf k}$ a typical phonon frequency of the system, which 
we take to be about 30 meV, independent of {\bf k}. 

\section{Phase diagram: some limiting cases} 
\label{pd_no_g:sec}

Preliminary to the full treatment of Sect.\ \ref{hf:sec}, 
we consider first the purely electronic problem in the absence of 
e-ph interaction. 
We start the discussion from particular limiting cases for which 
well-controlled statements, or at least intuitively clear ones, can be made,
without the need of any new specific calculations.
In the Appendix we will also consider, because it is useful in connection 
with the electron-phonon case, the unphysical limit of strong on-site 
attraction (large and negative $U$).

\subsection{Large positive $U$: the Mott insulator.} \label{large_u:sec}
For $U\gg V,W$, the system is deep inside the Mott insulating 
regime.\cite{Anderson_SE} The charge degrees of freedom are frozen,
with a gap of order U. The only dynamics is in the spin degrees of freedom.
Within the large manifold of spin degenerate states with exactly one electron 
per site, the kinetic energy generates, in second order perturbation theory, 
a Heisenberg spin-1/2 antiferromagnetic effective Hamiltonian
governing the {\it spin\/} degrees of freedom,
\begin{equation}
H_{\rm eff} = \sum_{(ij)} J_{ij} \, {\bf S}_{{\bf r}_i} \cdot
{\bf S}_{{\bf r}_j} \;,
\end{equation}
with $J_{ij}=4|t_{ij}|^2/U$.\cite{Anderson_SE}

For our test case of Si(111)/Si, the values of the hoppings are such that
$J_1 \approx 20$ meV, $J_2/J_1\approx 0.12$ while the remaining couplings 
$J_3,\cdots$ are very small.  
Antiferromagnetism is frustrated on the triangular lattice. 
Zero temperature long range order (LRO) -- if present -- should be of the 
three-sublattice $120^o$-N\'eel type, which can be also seen as a commensurate 
spiral spin density wave (s-SDW).

Because it does not imbalance charge, this state is not further affected by
electron-phonon coupling.

In summary, we expect for large values of $U$ a wide-gap Mott insulator with 
a s-SDW (spins lying in a plane, forming $120^o$ angles), 
a $3\times 3$ {\it magnetic} unit cell, but uniform charge (no CDW). 
This is, most likely, the state to be found on the Si-terminated and
C-terminated SiC(0001) surface at T=0 \cite{Northrup,Anisimov}.

\subsection{Strong inter-site repulsion:
an asymmetric CDW with three inequivalent sites.}
\label{large_v:sec}
The e-ph coupling can effectively reduce $U$, but not $V$. 
Therefore, it is of interest to consider the hypothetical regime
$W<U\ll V$.
When the first-neighbor electron-electron repulsion $V$ is large the
system, in order to minimize the interaction energy, will prefer a
$3\times 3$ CDW-like ground state, with two electrons on one sublattice (A), 
a single electron on another sublattice (B), and
zero electrons on the third sublattice (C) (see Fig.\ \ref{triang_cdw:fig}).
These states are still highly degenerate (in the absence of hopping) due to spin
degeneracy for the single unpaired electron on sublattice B.
A gap $U$ separates these states from the lowest-energy excited
configurations (see Fig.\ \ref{triang_cdw:fig}).
The spin degeneracy can be removed in second-order perturbation theory, 
owing to $t_2$, which leads to an effective spin-1/2 Heisenberg Hamiltonian 
within sublattice B, 
\begin{equation}
H_{\rm eff} = J \sum_{(ij)}^{\rm sublattice \, B}
{\bf S}_{{\bf r}_i} \cdot {\bf S}_{{\bf r}_j} \;,
\end{equation}
with a weak antiferromagnetic exchange constant $J=4t^2_2/U$.\cite{Anderson_SE}
Summarizing, we expect in this regime a strong $3\times 3$ asymmetric CDW 
(a-CDW) with three inequivalent sites ($\phi_{\rho}\approx \pi/6$, see below), 
and a spiral $3\sqrt{3}\times 3\sqrt{3}$ SDW, 
governing the unpaired electron spins, superimposed on it.  
Notice that, while the charge periodicity is $3\times 3$, the actual unit
cell is larger, i.e., $3\sqrt{3}\times 3\sqrt{3}$.
Despite having the correct charge periodicity, namely  $3\times 3$, this
a-CDW is not compatible with the experimental findings on Pb-Sn/Ge,
which is a symmetric CDW. 
We conclude that the low-temperature CDW state of these systems is not 
completely dominated by $V$.

\section{Mean-field theory.} \label{hf:sec}
In order to get a more complete picture of additional phases for smaller $U$, 
and of the possible phase diagram of the model we now turn to a quantitative 
mean field theory analysis.
The first issue is to include the possibility of magnetic correlations.
For small values of the interactions $U$ and $V$, the Stoner criterion can 
be used to study the possible magnetic instabilities of the 
paramagnetic metal obtained from LDA calculations.
The charge and spin susceptibilities are given, within the random phase
approximation,\cite{Mahan} by 
\begin{eqnarray}
\chi_C(\bf q) &\,=\,&
\frac{2 \chi_o({\bf q}) }{1 + (U+2V_{\bf q})\chi_o({\bf q})}
\nonumber \\
\chi_S(\bf q) &\,=\,& \frac{\chi_o({\bf q}) }{1 - U\chi_o({\bf q})}  \;,
\end{eqnarray}
where $\chi_o$ is the non-interacting susceptibility per spin projection,
and both factors of $2$ account for spin degeneracy.
The divergence of $\chi_S$ is governed, in this approximation, by $U$ only.
Since $\chi_o({\bf q})$ is finite everywhere, a finite $U$ is needed in
order to destabilize the paramagnetic metal. 
The wavevector ${\bf q}^*$ at which $\chi_S$ first diverges, by increasing $U$,
is in general incommensurate with the underlying unit cell.
The instability is towards an incommensurate, metallic, 
spiral SDW.\cite{KMurthy}
Fig.\ \ref{band_bz_chi0:fig} shows that, in our case, ${\bf q}^*=(1.32 K,0)$ 
(with $K=4\pi/3a$, the BZ corner). We get $U_c^{HF}/t_1\approx 3.7$.
(The other maximum of $\chi_o$ at ${\bf q}=(0.525 K,0)$
is very close to the result obtained for the triangular lattice with n.n.\
hopping only.\cite{KMurthy})
As for the charge susceptibility, a divergence can be caused only by
an attractive Fourier component of the potential $V_{\bf q}$. 
$V_{\bf q}$ has a minimum at the BZ corners $\pm {\bf K}$, with
$V_{\pm {\bf K}}=-3V$ for the n.n. model (A)
($V_{\pm {\bf K}}\approx -1.5422 V$ if a Coulomb tail is added, model B). 
This minimum leads to an instability towards a $3\times 3$ CDW as 
$(U+2V_{{\bf K}})\chi_o({\bf K})=-1$, i.e., given our value of
$\chi_o({\bf K})\approx 0.2/t_1$,
$(U+2V_{{\bf K}})\approx -5t_1$.
For model A we get a transition, when $U=0$, at $V_c^{MF}/t_1\approx 0.83$.

In general, the small coupling paramagnetic metal is surrounded by an 
intermediate coupling region, where complicated incommensurate --
generally metallic -- solutions occur. 
For stronger $U$ and $V$, commensurate solutions are privileged.\cite{KMurthy}
In view of the fact that a $3\times 3$ CDW is experimentally relevant, we
concentrate our analysis on the simplest commensurate phases. 
These are easy to study with a standard Hartree-Fock (HF) mean-field 
theory.
In particular, we restrict ourselves to order parameters
associated with non-vanishing momentum space averages of the type
$\langle c^{\dagger}_{{\bf k},\sigma} c_{{\bf k},\sigma '} \rangle$ and 
$\langle c^{\dagger}_{{\bf k},\sigma} c_{{\bf k}\pm{\bf K},\sigma '}\rangle$.  
Possible non-vanishing order parameters are the uniform magnetization 
density ${\bf m}$, 
\begin{equation} \label{m:eqn}
{\bf m} = \frac{1}{N_s} \sum_{\bf k}^{BZ} \sum_{\alpha,\beta}
\langle c^{\dagger}_{{\bf k},\alpha} ({\vec \sigma})_{\alpha\beta}
c_{{\bf k},\beta} \rangle  = \frac{2}{N_s} \langle {\bf S}_{\rm tot} \rangle \;,
\end{equation}
the ${\bf K}$-component of the charge density,
\begin{equation} \label{rho:eqn}
\rho_{\bf K} = \frac{1}{N_s} \sum_{\bf k}^{BZ} \sum_{\sigma}
\langle c^{\dagger}_{{\bf k},\sigma}
c_{{\bf k}-{\bf K},\sigma}  \rangle \;,
\end{equation}
and the ${\bf K}$-component of the spin density
\begin{equation}
{\bf S}_{\bf K} = \frac{1}{N_s} \sum_{\bf k}^{BZ} \sum_{\alpha,\beta}
\langle c^{\dagger}_{{\bf k},\alpha} \frac{({\vec \sigma})_{\alpha\beta}}{2}
c_{{\bf k}-{\bf K},\beta} \rangle \;.
\end{equation}
Note that only $\rho_{\bf K}$ and ${\bf S}_{\bf K}$ are $3\times 3$ periodic.
Moreover, $K$-components of bond order parameters of the type
$\langle c^{\dagger}_{{\bf r},\sigma} c_{{\bf r}',\sigma '} \rangle$ are
automatically included in the calculation.
$\rho_{\bf K}$ and ${\bf S}_{\bf K}$ have phase freedom, and 
are generally complex: $\rho_{\bf K}=|\rho_{\bf K}| e^{i\phi_{\rho}}$, etc.
The role of the phase is clarified by looking at the real-space 
distribution within the $3\times 3$ unit cell. 
For the charge, for instance, 
$\langle n_{{\bf r}_j}\rangle=1+2|\rho_{\bf K}|\cos{(2\pi p_j/3+\phi_{\rho})}$,
where $p_j=0,1,2$, respectively, on sublattice A, B, and C. 
The e-ph coupling is included but, after linearization, the
displacement order parameter is not independent, and is given by
$\langle z_{\bf K}\rangle=(g/M\omega_{\bf K}^2)\rho_{\bf K}$.
Only the phonon modes at $\pm{\bf K}$ couple directly to the CDW. 
The phonon part of the Hamiltonian can be diagonalized by displacing
the oscillators at $\pm{\bf K}$. 
This gives just an extra term in the electronic HF Hamiltonian of the form
$\Delta U (\rho^*_{\bf K} {\hat \rho}_{\bf K} + {\rm H.c.})$, 
with an energy $\Delta U=-g^2/M\omega^2_{\bf K}$ which is the relevant 
coupling parameter.  
This term acts, effectively, as a negative-$U$ contribution acting only
on the charge part of the electronic Hamiltonian.

With the previous choice of non-vanishing momentum space averages, the 
Hartree-Fock Hamiltonian reads:
\begin{eqnarray} \label{hf_ham:eqn}
H_{\rm H-F} &\,=\,&
\sum_{{\bf k}}^{BZ} \sum_{\sigma} \epsilon_{\bf k} n_{{\bf k},\sigma}
\,-\, U \, {\bf m} \cdot {\bf S}_{\rm tot} \nonumber \\
&& + \sum_{\bf k}^{BZ} \sum_{\sigma}
\left\{ \left[ \left( \frac{U}{2} + V_{\bf K} 
               -\frac{g^2}{M\omega^2_{\bf K}} \right) \rho_{\bf K}
- \sigma U S^z_{\bf K} \right]
c^{\dagger}_{{\bf k},\sigma} c_{{\bf k}+{\bf K},\sigma} + {\rm H.c.} \right\}
\nonumber \\
&& - U \sum_{\bf k}^{BZ} \left\{
S^+_{\bf K} c^{\dagger}_{{\bf k},\downarrow} c_{{\bf k}+{\bf K},\uparrow} \,+\,
S^-_{\bf K} c^{\dagger}_{{\bf k},\uparrow} c_{{\bf k}+{\bf K},\downarrow}
\,+\, {\rm H.c.} \right\} \nonumber \\
&& + \sum_{\bf k}^{BZ} \sum_{\sigma}
\left\{
A^{(\sigma\sigma)}_{\bf k}
c^{\dagger}_{{\bf k},\sigma} c_{{\bf k},\sigma}
\,+\, \left[ B^{(\sigma\sigma)}_{\bf k}
c^{\dagger}_{{\bf k},\sigma} c_{{\bf k}+{\bf K},\sigma} + {\rm H.c.} \right]
\right\} \nonumber \\
&& + \sum_{\bf k}^{BZ} \sum_{\sigma}
\left\{
A^{({\bar \sigma}\sigma)}_{\bf k}
c^{\dagger}_{{\bf k},{\bar \sigma}} c_{{\bf k},\sigma}
\,+\, \left[ B^{({\bar \sigma}\sigma)}_{\bf k}
c^{\dagger}_{{\bf k},{\bar \sigma}} c_{{\bf k}+{\bf K},\sigma} + {\rm H.c.}
\right] \right\} \;.
\end{eqnarray}
The last two terms originate exchange contributions due to the $V$-term;
$A^{(\sigma'\sigma)}_{\bf k}$ and $B^{(\sigma'\sigma)}_{\bf k}$ are shorthands
for the following convolutions:
\begin{eqnarray} \label{AB:eqn}
A^{(\sigma'\sigma)}_{\bf k} &\,=\,& - \frac{1}{N_s} \sum_{{\bf k}'}^{BZ}
V_{{\bf k}-{\bf k}'} \,
\langle c^{\dagger}_{{\bf k}',\sigma} c_{{\bf k}',\sigma'} \rangle \nonumber \\
B^{(\sigma'\sigma)}_{\bf k} &\,=\,& - \frac{1}{N_s} \sum_{{\bf k}'}^{BZ}
V_{{\bf k}-{\bf k}'} \,
\langle c^{\dagger}_{{\bf k}'+{\bf K},\sigma} c_{{\bf k}',\sigma'} \rangle \;.
\end{eqnarray}

The BZ is divided into three regions: a reduced BZ (RBZ), and the two
zones obtained by ${\bf k}\pm{\bf K}$ with ${\bf k}\in$ RBZ.
The HF problem in Eq.\ \ref{hf_ham:eqn} reduces to the self-consistent 
diagonalization of a $6\times 6$ (including the spin) matrix for each 
${\bf k}\in$ RBZ.

\subsection{Landau theory} \label{Landau:sec}

The mean-field solutions must be compatible with the symmetry of the
problem.
Before discussing the HF phase diagram we obtain, it is useful
to present a few general phenomenological considerations based on a symmetry
analysis of the Landau theory built from the CDW order parameter
$\rho_{\bf K}$ (a complex scalar), the SDW order parameter ${\bf S}_{\bf K}$
(a complex vector), and the uniform magnetization ${\bf m}$
(a real vector).\cite{Toledano}
In the absence of spin-orbit coupling, the possible contributions to the
Laundau free energy $F$ allowed by symmetry, up to fourth order, have the form
\begin{eqnarray}
F &\,=\,& \frac{1}{2} a_{\rho} |\rho_{\bf K}|^2
    \,+\, \frac{1}{2} a_{m}    |{\bf m}|^2
    \,+\, \frac{1}{2} a_{s}    |{\bf S}_{\bf K}|^2
    \,+\, F_{3} \,+\, F_{4} \nonumber \\
F_3 &\,=\,& ( B_{\rho} \rho_{\bf K}^3 + {\rm c.c.} )
  \,+\, [ B_{\rho s} \rho_{\bf K} ({\bf S}_{\bf K} \cdot {\bf S}_{\bf K} )
          \,+\, {\rm c.c.} ] \nonumber \\
F_4 &\,=\,& b_{\rho} |\rho_{\bf K}|^4 + b_{m} |{\bf m}|^4
   \,+\, b_s^{(1)} |{\bf S}_{\bf K}|^4
   \,+\, b_s^{(2)} ({\bf S}_{\bf K} \times {\bf S}^*_{\bf K})^2 \nonumber \\
&& +\, b_{\rho s} |\rho_{\bf K}|^2 |{\bf S}_{\bf K}|^2
   +\, b_{\rho m} |\rho_{\bf K}|^2 |{\bf m}|^2
   +\, b_{m s}^{(1)} |{\bf m}|^2 |{\bf S}_{\bf K}|^2
   +\, b_{m s}^{(2)} ({\bf m} \cdot {\bf S}_{\bf K})
                     ({\bf m} \cdot {\bf S}^*_{\bf K}) \nonumber \\
&& +\, [b_{m s}^{(3)} ({\bf m} \cdot {\bf S}_{\bf K})
                      ({\bf S}_{\bf K} \cdot {\bf S}_{\bf K}) + {\rm c.c.}]
 \,+\, [b_{\rho m s} \rho^2_{\bf K}
                     ({\bf m} \cdot {\bf S}_{\bf K}) + {\rm c.c.}] \;,
\end{eqnarray}
with $|{\bf S}_{\bf K}|^2=({\bf S}_{\bf K} \cdot {\bf S}^*_{\bf K})$.
Notice that third order invariants are present due to commensurability,
$3{\bf K}={\bf G}$ (reciprocal lattice vector). Therefore, first order
transitions are generally possible.\cite{Toledano}
 
This expansion suggests a number of additional comments:
{\it i\/}) A CDW can occur without accompanying magnetism, i.e.,  
$\rho_{\bf K}\ne 0$, while ${\bf m}=0$ and ${\bf S}_{\bf K}=0$.
This is the case, as we shall see later, for the small $U$ region of the HF
phase diagram.
{\it ii\/}) The possible SDW phases are either collinear (l-SDW)
(for which $({\bf S}_{\bf K} \times {\bf S}^*_{\bf K})=0$) or
coplanar.\cite{Kivelson}
The latter have, with a suitable choice of the phases,
${\bf S}^x_{\bf K}=|{\bf S}_{\bf K}|\cos{\alpha}$ and
${\bf S}^y_{\bf K}=-i |{\bf S}_{\bf K}|\sin{\alpha}$,
and can be generally described as a spiral SDW (s-SDW)
\begin{equation}
\langle {\bf S}_{\bf r} \rangle \,=\, 2 |{\bf S}_{\bf K}| \,
[ {\hat{\bf x}} \cos{\alpha} \, \cos{({\bf K}\cdot {\bf r})} -
  {\hat{\bf y}} \sin{\alpha} \, \sin{({\bf K}\cdot {\bf r})} ] \;,
\end{equation}
with an eccentricity parameter $\alpha\ne 0,\pi/2$.
($\alpha=0$ or $\pi/2$ are actually l-SDW along the ${\hat{\bf x}}$
or ${\hat{\bf y}}$ directions.)
$\alpha=\pi/4$ describes a circular spiral SDW.
Now, the only possibility of having a SDW without CDW is via a circular spiral
SDW ($\alpha=\pi/4$). Indeed, the third order invariant
$[B_{\rho s} \rho_{\bf K} ({\bf S}_{\bf K} \cdot {\bf S}_{\bf K})+{\rm c.c.}]$
vanishes by symmetry only for a circular spiral SDW, for which
$({\bf S}_{\bf K} \cdot {\bf S}_{\bf K})=0$; in all other cases, a SDW
implies -- if $B_{\rho s}\ne 0$ -- a CDW as well.
{\it iii\/}) The simultaneous presence of a SDW and a CDW implies, generally, 
a finite magnetization $\bf m$, via the fourth order invariant
$[b_{\rho m s} \rho^2_{\bf K} ({\bf m} \cdot {\bf S}_{\bf K})+{\rm c.c.}]$,
unless the phases of $\rho$ and $S$ are such that 
$2\phi_{\rho}+\phi_{\sigma}=\pi/2+n\pi$. 
This happens in phase E of our phase diagram, which has therefore no 
uniform magnetization.
{\it iv\/}) The presence of a SDW leads, generally, to a finite uniform
magnetization as well, via the fourth order invariant
$[b_{m s}^{(3)} ({\bf m} \cdot {\bf S}_{\bf K})
({\bf S}_{\bf K} \cdot {\bf S}_{\bf K}) + {\rm c.c.}]$,
unless the phase $\phi_{\sigma}$ is such that
$3\phi_{\sigma}=\pi/2+m\pi$.

\subsection{Phase diagram in the Hartree-Fock approximation} \label{HF_pd:sec}

We present a brief summary of the mean-field HF calculations for 
arbitrary $U$, $V$, and $g$, obtained by solving numerically
the self-consistent problem in Eqs.\ \ref{m:eqn}-\ref{AB:eqn}.
The main phases present in the HF phase diagram are shown in 
Fig.\ \ref{hf_pd:fig} for the case of $g=0$. 
The effect of $g\ne 0$ will be discussed further below.
  
{\bf Phase A: Spiral SDW insulating phase.}
The circular spiral SDW (phase A) dominates the large $U$, small $V$ part of 
the phase diagram, as expected from the Heisenberg model mapping at 
$U\to\infty$ (see sect.\ \ref{large_u:sec}). 
This is the Mott insulator phase, probably relevant for SiC.
Its HF bands are shown in Fig.\ \ref{hf_bands:fig}(a).

{\bf Phase A': Collinear SDW with $m^z=1/3$ insulating phase.}
This is another solution of the HF equations in the large $U$, 
small $V$ region. 
It is an insulating state characterized by a linear l-SDW plus a small 
CDW with $\phi_{\rho}=0$, accompanied by a magnetization $m^z=1/3$ (phase A').
This collinear state lies above the s-SDW by only a small energy difference 
(of order $0.03 t_1$ per site), and could be stabilized by other factors
(e.g., spin-orbit). 
A recent LSDA calculation for $\sqrt{3}$-Si/Si(111) has indicated this
l-SDW as the ground state, at least if spins are forced to be 
collinear.\cite{Sandro}
The HF bands for this solution are shown in Fig.\ \ref{hf_bands:fig}(b), 
and are very similar to the LSDA surface band for Si/Si(111). 
The phase $\phi_{\rho}=0$ of the CDW order parameter corresponds 
to a real-space charge distribution in which one sublattice has a charge 
$1+2|\rho_{\bf K}|$, while the remaining two are equivalent and have 
charges $1-|\rho_{\bf K}|$, compatible with the experimental findings on 
Sn/Ge(111) and Pb/Ge(111). 
The amplitude $|\rho_{\bf K}|$ of the CDW is in general quite small in this 
phase. 
It should be noted, however, that a STM map is not simply a direct measure 
of the total charge density.\cite{Selloni,Tosatti:high} 
This will be discussed in sect.\ \ref{stm:sec}.

{\bf Phase B': Asymmetric CDW with $m^z=1/3$ insulating phase.}
By increasing the n.n.\ repulsion $V$, the energies of the s-SDW 
and of the l-SDW tend to approach, 
until they cross at a critical value $V_c$ of $V$. 
At $U/t_1=10$ we find $V_c/t_1\approx 3.3$ for model A, $V_c/t_1\approx 6.6$ 
for model B.
As $V>V_c$, however, an insulating asymmetric CDW (a-CDW) prevails. 
This is simply the spin collinear version of the non-collinear phase
described in Sect.\ \ref{large_v:sec}. 
Fig.\ \ref{e_cdw:fig} shows the energy per site of the most relevant
HF solutions at $U/t_1=10$ as a function of $V$ for model B (Coulomb tail case).
The s-SDW and the l-SDW cross at $V_c\approx 6.6 t_1$ where,
however, the a-CDW insulating solution starts to be the favored one.
This large-$V$ solution has a large $CDW$ order parameter
with $\phi_{\rho}\ne 0$ (mod. $2\pi/3$), a concomitant l-SDW, and $m^z=1/3$.
By recalling the discussion in sect.\ \ref{large_v:sec}, we notice that
a state with a magnetization $m^z=1/3$ and a l-SDW is the best HF solution once
a $3\times 3$ restriction has been applied, since a spiral SDW on the
singly occupied sublattice would involve a larger periodicity (phase B).

{\bf Phase D: Symmetric non-magnetic CDW metallic phase.}
For small values of $U$ and $V$, or for large enough e-ph coupling
$g$, a {\em metallic\/} CDW with $\phi_{\rho}=0$ (m-CDW) is found. 
(See Fig.\ \ref{hf_bands:fig}(c) for the HF bands.)
This phase constitutes a candidate, alternative to the magnetic phase B', 
and compatible with the main experimental facts, which 
might be relevant for the case of Pb/Ge(111) and of Sn/Ge(111). 
The degree of metallicity of this phase is much reduced relative to the 
undistorted surface (pseudo-gap). 

We stress that the e-ph interaction can stabilize the 
$\phi_{\rho}=0$ m-CDW also at relatively large $U$, by countering $U$ with a
large negative $\Delta U=-g^2/M\omega^2_{\bf K}$.  
We demonstrate this in Fig.\ \ref{e_u8v2_ph:fig}, where we plot the energy
per site as a function of $\Delta U$ at $U/t_1=8$ and $V/t_1=2$, for the 
three relevant HF solutions, i.e., the spiral SDW (phase A), the collinear
SDW with $m^z=1/3$ (phase A'), and the metallic non-magnetic CDW (phase D). 
The spiral SDW is unaffected by the electron-phonon coupling. 
The energy of the collinear SDW with $m^z=1/3$ improves a little bit 
by increasing $g$, due to the small CDW amplitude of this phase. This effect
is not large enough as to make this phase stable in any range of couplings.
At a critical value of $g$, the metallic non-magnetic CDW  
(where the CDW order parameter is large, $|\rho_{\bf K}| \sim 0.5$) wins over
the magnetic phases. 
The Fourier transform of the lattice distortion at ${\bf K}$ is given by 
$\langle z_{\bf K}\rangle=(g/M\omega_{\bf K}^2)\rho_{\bf K}
=\rho_{bf K} |\Delta U|/g$.

A rough estimate shows that the order of magnitude
of the electron-phonon coupling necessary to stabilize the CDW phase is not
unreasonable.
With $g=1$ eV/$\AA$, $M_{Si}=28$, and $\omega_{\bf K}\approx 30$ meV
we get $\Delta U \approx -3 t_1$, sufficient to switch from a 
s-SDW ground state to a m-CDW for $U/t_1=8$ and $V/t_1=2$. 
With these values of the parameters we have $|\rho_{\bf K}| \approx 0.43$, and
we estimate $|\langle z_{\bf K}\rangle| \approx 0.07 \AA$. This corresponds,
since $\langle z_{\bf r} \rangle \sim 2\cos({\bf K}\cdot {\bf r}) 
|\langle z_{\bf K}\rangle|$, to a total displacement between the adatom
going up and the two going down of $\Delta z = 3
|\langle z_{\bf K}\rangle| \approx 0.2\AA$. 

We notice that values of $g$ much larger than those used in 
Fig.\ \ref{e_u8v2_ph:fig} would eventually stabilize a superconducting 
ground state (see Appendix).

\section{CDW order parameter and STM experiments} \label{stm:sec}

We discuss, in the present section, the relationship between the
CDW order parameter, as defined in Eq.\ \ref{rho:eqn}, and an STM map
of the surface. 
As the crudest approximation to the tunneling current for a given bias 
$V_{\rm bias}$ we consider the integral of the charge density for 
one-electron states within $V_{\rm bias}$ from the Fermi level, 
weighted with barrier tunneling factor $T(V)$,\cite{Selloni,Tosatti:high}
\begin{equation}
\label{stm:eqn}
J(V_{\rm bias},{\bf r}=x,y;z) \approx \int_0^{V_{\rm bias}} dV  \sum_{n{\bf k}} 
|\Psi_{n{\bf k}}({\bf r})|^2 \delta (E_{n{\bf k}}-E_F+V) T(V)  \;.
\end{equation}
The tunneling factor leads to weighting prominently the states
immediately close to the Fermi level. 
In view of the purely qualitative value of Eq.\ (\ref{stm:eqn}), 
we have moreover decided to ignore $T(V)$ altogether and to account for 
its effect by reducing the bias voltage
$V_{\rm eff}$ in Eq.\ (\ref{stm:eqn}), to an effective value 
$V_{\rm bias}^{\rm eff}$. 
By doing this, we have extracted an ``STM map'' for a point in phase A'
($U/t_1=9$ and $V=2$, model A)
-- a spin-density waves where the amplitude of the CDW order parameter is 
rather small, $|\rho_{\bf K}|=0.039$ --
and a point in phase D ($U/t_1=4$ and $V=2$, model A) 
-- a pure CDW where the order parameter is quite large,
$|\rho_{\bf K}|=0.4$. 
The results for constant $z$, and $x,y$ moving from adatom A to B to C,
are shown in Fig.\ \ref{stm:fig}(a) and (b), for the two cases. 
The solid curves refer to positive bias (current flowing from the sample
to the tip), probing occupied states close to the Fermi level. 
The dashed curve refers to negative bias, probing unoccupied states.
In both cases a) and b), one of the three atoms yields a larger current at 
positive bias, while the other two atoms have larger currents at negative bias. 
The insets show the predicted ``contrast'' between the two peak values,
$(J_1-J_2)/(J_1+J_2)$, $J_1$ and $J_2$ being in each case, respectively, 
the largest and the smallest of the STM peak currents at the positions of 
the adatoms.
We notice the following points: 
i) for the occupied states (positive bias) the pure CDW phase has, as expected,
a larger contrast than the magnetic phase. 
As we neglect the tunneling factors $T(V)$, in the limit of large positive
effective bias we recover the total asymmetry in the charge of the two 
inequivalent atoms, $(n_1-n_2)/(n_1+n_2)$, indicated by a dashed horizontal 
line in the insets.  
Observe that the way this large bias limit is reached is completely different 
for the two cases a) and b): in the magnetic case a) the contrast overshoots 
at small biases attaining values substantially larger than the nominal CDW order
parameter, and then goes to the limit $(n_1-n_2)/(n_1+n_2)$ from above;
in the pure CDW case b), on the contrary, the limit is reached monotonically
from below. 
ii) for empty states (negative bias) the contrast is even more surprising: at
small bias it is very large in both cases a) and b). By increasing
the bias, the contrast for the pure CDW case tends monotonically to a large 
value, whereas the magnetic case shows a strong non monotonicity. 

These results suggest that one should look more carefully, and quantitatively,
at the behavior of the asymmetry between STM peak currents as a function
of the bias, including the region of relatively small biases: the different
behavior of the asymmetry of the magnetic case versus the pure CDW case should 
be marked enough -- and survive in a more refined analysis including $T(V)$ --
as to make the STM map a good way of discriminating between the two scenarios. 

\section{Discussion and conclusions} \label{discussion:sec}

Within our model study we have learned that on the surfaces considered: 

(i) If $U$ and $V$ are ignored, there is no straight electron-phonon driven  
$3\times 3$ CDW surface instability. 
However, any phase involving a CDW, for example as
a secondary order parameter attached to a primary SDW, can take
advantage and gain some extra stabilization energy from a small surface
lattice distortion, via electron-phonon coupling.

(ii) Electron-electron repulsion and the two-dimensional Fermi Surface
are capable of driving transitions of the undistorted metallic surface 
to a variety of states, that are either insulating or in any case less 
metallic, some possessing the $3\times 3$ periodicity. 

(iii) This can occur via two different mechanisms: a) the inter-site
repulsion $V$ can stabilize insulating or semi-metallic CDWs, without 
a crucial involvement of spin degrees of freedom; b) the on-site 
repulsion $U$ can produce essentially magnetic insulators with or without 
a weak accompanying $3\times 3$ CDW, as required by symmetry.

(iv) For $U$ moderate of order $W$ and for smaller $V$, an interesting state 
is realized, with a large SDW and a small accompanying CDW. The state 
is either a small-gap insulator, or a semi-metal, and may or may not 
be associated with a net overall magnetization, depending on the nature 
(linear or spiral, respectively) of the leading SDW. 

(v) For $U$ and $V$ both small but finite, a metallic CDW without any magnetism
is obtained. The same phase can also be stabilized for larger values of $U$
by the presence of a substantial electron-phonon coupling. 
We stress that, in this case, $V$ is the coupling responsible for the
$3\times 3$ symmetry of the unit cell, whereas the role of the electron-phonon 
coupling is that of destroying magnetism by effectively decreasing $U$.
Electron-phonon coupling alone is not sufficient to justify a commensurate
$3\times 3$ CDW.

(vi) Either of the phases in (iv) or (v) could be natural candidates for 
explaining the weak $3\times 3$ CDW seen experimentally on Sn-Pb/Ge(111). 

(vii) Finally, for large $U$, small $V$ (in comparison with the bandwidth $W$)
the  Mott-Hubbard state prevails. It is a wide-gap insulator, with
a pure spiral SDW, with $3\times 3$ overall periodicity, and coplanar 
$120^o$ long-range spin ordering at zero temperature. It possesses
no net magnetization, and no accompanying CDW. 

(viii) The above is the kind of state which we expect to be realized on 
SiC(0001), and also possibly on K/Si(111):B. 

Among existing experiments, we have addressed particularly
photoemission and STM. Our calculated band structure for both the SDW/CDW
state A' (iv) and the pure CDW state D (v) exhibit features which are similar 
to those found in photoemission from 
Sn-Pb/Ge(111).\cite{Modesti_2,Avila,LeLay,asensio}  
The simulated STM images for the two kind of states are predicted to differ
in their voltage dependence. 

Future experiments are strongly called for, aimed at detecting whether
magnetic correlations are actually dominant, as we think is very likely, 
on all these surfaces, or whether Sn-Pb/Ge(111) are instead non-magnetic
and electron-phonon driven. 
The issue of whether magnetic long-range order -- which we definitely propose
for SiC(0001) and K/Si(111):B at $T=0$, and also hypothesize for 
Sn-Pb/Ge(111) -- survives up to finite temperatures is one which we 
cannot settle at this moment. This due to the difficulty
in estimating the surface magnetic anisotropy, without which order
would of course be washed out by temperature. In any case, it should be
possible to pursue the possibility of either magnetism or incipient
magnetism using the appropriate spectroscopic tools.

This line of experimental research, although undoubtedly difficult, should 
be very exciting since it might lead to the unprecedented discovery of 
magnetic states at surfaces possessing no transition metal ions of any kind, 
such as these seemingly innocent semiconductor surfaces.  

We acknowledge financial support from INFM, through projects LOTUS and
HTSC, and from EU, through ERBCHRXCT940438. 
We thank S. Modesti, J. Lorenzana, M.C. Asensio, J. Avila, G. Le Lay, 
E.W. Plummer and his collaborators, for discussions.

\section{Appendix. Large negative $U$: a superconducting ground state.}
\label{neg_u:sec}
%
The limit of large negative $U$, $U\to -\infty$, is considered here
to show that CDWs are not favored by on-site attraction alone. Instead,
a superconducting ground state is favored.\cite{santos}
To see this, consider the real-space states which are the low energy
configurations for $U\to -\infty$: they consist of $N_e/2$ sites (if $N_e$
is the number of electrons) each of which is occupied by a pair of electrons
with opposite spins.
The large degeneracy in this manifold of states is
-- once again, like in the $U\to \infty$ case -- removed by kinetic energy
in second order perturbation theory. By assigning a pseudo-spin-1/2 state
to each site (up, if occupied by a pair, down if empty) one can show that
the effective Hamiltonian is \cite{santos}
\begin{equation}
H_{\rm eff} = -\sum_{(ij)} \frac{J^{\perp}_{ij}}{2} \,
\left( S^+_{{\bf r}_i} S^-_{{\bf r}_j} \,+\, {\rm H.c.} \right) \,+\,
\sum_{(ij)} J^{z}_{ij} \, S^z_{{\bf r}_i} S^z_{{\bf r}_j} \;,
\end{equation}
with $J^{\perp}_{ij}=4|t_{ij}|^2/|U|$ and $J^{z}_{ij}=J^{\perp}_{ij}$.
If $V$-terms are added, $J^z$ is modified to
$J^{z}_{ij}=J^{\perp}_{ij}+4V_{ij}$.
Restricting our consideration to the n.n. case, we are left with a
n.n.\ Heisenberg Hamiltonian with ferromagnetic xy-part and an 
antiferromagnetic z-part. The sign of the xy-part cannot be changed at will
by a canonical transformation because the lattice is non-bipartite. 
The result is that the order is in the plane (i.e., superconductivity wins)
for small $V$. Only if $V$ is large enough the CDW 
(i.e., order in the z-direction) will be favored. 

Entirely similar considerations apply to the case of strong electron-phonon
coupling, $g\to \infty$.


\newpage
\begin{center} {\bf Figure Captions} \end{center}

\begin{figure}
\caption{
Surface state dispersion for hypothetical Si/Si(111), as obtained from gradient
corrected LDA (solid squares). 
A very similar band is obtained for Pb/Ge(111) and Sn/Ge(111). 
The solid line is a tight-binding 
fit obtained by including up to the sixth shell of neighbors, $t_1,\cdots,t_6$.
The fit gives $t_1=0.0551$ eV, and $t_2/t_1=-0.3494$, $t_3/t_1=0.1335$,
$t_4/t_1=-0.0615$, $t_5/t_1=0.0042$, $t_6/t_1=-0.0215$.
The dashed line is the best fit using $t_1$ and $t_2$ only. 
Upper inset: The Fermi surface of the half-filled surface band. The outer 
hexagon is the BZ of the $\sqrt{3}\times\sqrt{3}$ phase, and the inner
hexagon is the BZ of the $3\times 3$ phase. 
Notice the poor nesting at the BZ corner wavevector ${\bf K}=(4\pi/3a,0)$,
joining two opposite $M_{3\times 3}$ points.
Lower inset: The zero temperature Lindhard response function $\chi_o(\bf q)$
for the half-filled surface band.
Notice the two peaks located at ${\bf q}_1\approx 0.525 {\bf K}$ and
${\bf q}_2\approx 1.32 {\bf K}$, and no feature whatsoever at ${\bf K}$,
indicating poor nesting.
}
\label{band_bz_chi0:fig}
\end{figure}
%
\begin{figure}
\caption{ Density contours of the Wannier function associated with
         the Si/Si(111) surface band, calculated with gradient-corrected
	 LDA: dots correspond to atomic positions.
}
\label{wannier:fig}
\end{figure}
%
\begin{figure}
\caption{The CDW on the adatom triangular lattice in the limit $t_{ij}=0$,
$U\ll V$. Sublattice A is doubly occupied, sublattice B is singly occupied,
and C is empty. The large spin degeneracy associated to the unpaired singly
occupied sites on sublattice B is removed by the next-nearest neighbor hopping
$t_2$ in second order perturbation theory.
The arrow indicates a possible virtual process, leading to an extra doubly
occupied site (with an associated energy gap of $U$), which generates the
standard antiferromagnetic exchange coupling between the spins on sublattice B.
}
\label{triang_cdw:fig}
\end{figure}
%
\begin{figure}
\caption{
Schematic Hartree-Fock phase diagram of model A (nearest neighbor
$V$ only, $g=0$) for the band structure shown in 
Fig.\ \protect\ref{band_bz_chi0:fig}.
Only the most important commensurate $3\times 3$ phases have been studied.
(Details of the merging of the various lines are not accurate.)
Left figure: phase obtained allowing spin non-collinearity. 
Right figure: strictly collinear phase diagram. 
The non-vanishing order parameters of the different phases are as follows:
A: ${\bf S}_{\bf K}=|{\bf S}_{\bf K}|/\sqrt{2}(1,-i,0)$; 
A': ${\bf S}_{\bf K}=|{\bf S}_{\bf K}|(0,0,1)$, $m_z=1/3$, 
$\rho_{\bf K}=|\rho_{\bf K}|$; 
B': ${\bf S}_{\bf K}=|{\bf S}_{\bf K}|(0,0,1)$, $m_z=1/3$, 
$\rho_{\bf K}=|\rho_{\bf K}|e^{i\phi_{\rho}}$ (with $0<\phi_{\rho}<\pi/6$); 
B: $3\sqrt{3}$ extension of phase B' when allowing for non-collinearity
of the unpaired spins.
C and C' are semi-metallic versions of A and A', respectively.
D: $\rho_{\bf K}=|\rho_{\bf K}|$; 
E: ${\bf S}_{\bf K}=i|{\bf S}_{\bf K}|(0,0,1)$, 
$\rho_{\bf K}=-|\rho_{\bf K}|$; 
F: $\rho_{\bf K}=|\rho_{\bf K}|e^{i\phi_{\rho}}$ (with $0<\phi_{\rho}<\pi/6$); 
IM: Incommensurate metallic SDW/CDW; 
PM: Paramagnetic metal.
Phases A(A'), and B(B') are insulating and magnetic. 
Phases C(C'), E, and IM are metallic and magnetic. 
Phase D is a pure CDW and is metallic.
Phases A', C', and D have CDW order parameter with the same symmetry as 
observed on Pb/Ge(111) and Sn/Ge(111). 
The effect of a finite electron-phonon coupling ($g\ne 0$) is discussed 
in the text.
}
\label{hf_pd:fig}
\end{figure}
%
\begin{figure}
\caption{
Plot of the HF electronic bands along high symmetry directions of the BZ
for the s-SDW and two CDW $\phi_{\rho}=0$ solutions:
(a) at $U/t_1=9$ and $V/t_1=2$, the insulating s-SDW (phase A, ground state);
(b) at $U/t_1=9$ and $V/t_1=2$, the insulating solution with a small
CDW and $m^z=1/3$ (phase A', meta-stable, the actual ground state being 
the s-SDW);
Solid and dashed lines denote up and down bands, respectively.
(c) at $U/t_1=4$ and $V/t_1=2$, the metallic solution with a large
CDW and no magnetism (phase D).
The band structure for phases C and C' are similar to A and A' except for
band overlap, making them semi-metallic. 
Insets indicate the charge and spin imbalance (when present) between the 
three adatoms in the $3\times 3$ unit cell.  
}
\label{hf_bands:fig}
\end{figure}
%
\begin{figure}
\caption{Energy per particle, as a function of $V$ at $U/t_1=10$,
for the commensurate HF solutions of phase A, A', and B', 
obtained for model B (Coulomb tail case).
The results for model A are entirely similar, with $V_c\approx 3.3t_1$.
}
\label{e_cdw:fig}
\end{figure}
%
\begin{figure}
\caption{The energy per site, as a function of the electron-phonon 
coupling $-\Delta U=g^2/M\omega^2_{\bf K}$, at $U=8t_1$ and $V=2t_1$ for the
model with n.n. interactions only, for the three relevant HF solutions, i.e.,
the spiral SDW, the collinear SDW with $m^z=1/3$, and the metallic non-magnetic
CDW. 
}
\label{e_u8v2_ph:fig}
\end{figure}
%
\begin{figure}
\caption{(a) STM map for tip motion (at constant height) along the triangle 
indicated, for the l-SDW state at $U/t_1=9$, $V=2$. The result is
obtained from Eq.\ \protect\ref{stm:eqn} without explicitly including $T(V)$, 
but reducing the bias to an effective one, $V_{\rm bias}^{\rm eff}=6t_1$, 
thus imposing that only states sufficiently close to the Fermi level 
contribute significantly to the STM current. The solid and dashed curves 
refer, respectively, to positive and negative STM bias.
The inset shows the ``contrast'' between the peaks (see text) as a function
of the effective bias $V_{\rm bias}^{\rm eff}$.
(b) Same as (a) for a pure CDW state in phase D.
}
\label{stm:fig}
\end{figure}


\begin{thebibliography}{99}
\bibitem{non-diag:nota}
We observe that, in principle, non-diagonal terms, i.e., terms which
cannot be recast in the form of a density-density interaction, should
be included. However, the magnitude of such terms -- which involve overlap
integrals between Wannier orbitals at different adatoms -- can be estimated
to be quite smaller than the diagonal terms we keep.
\end{thebibliography}
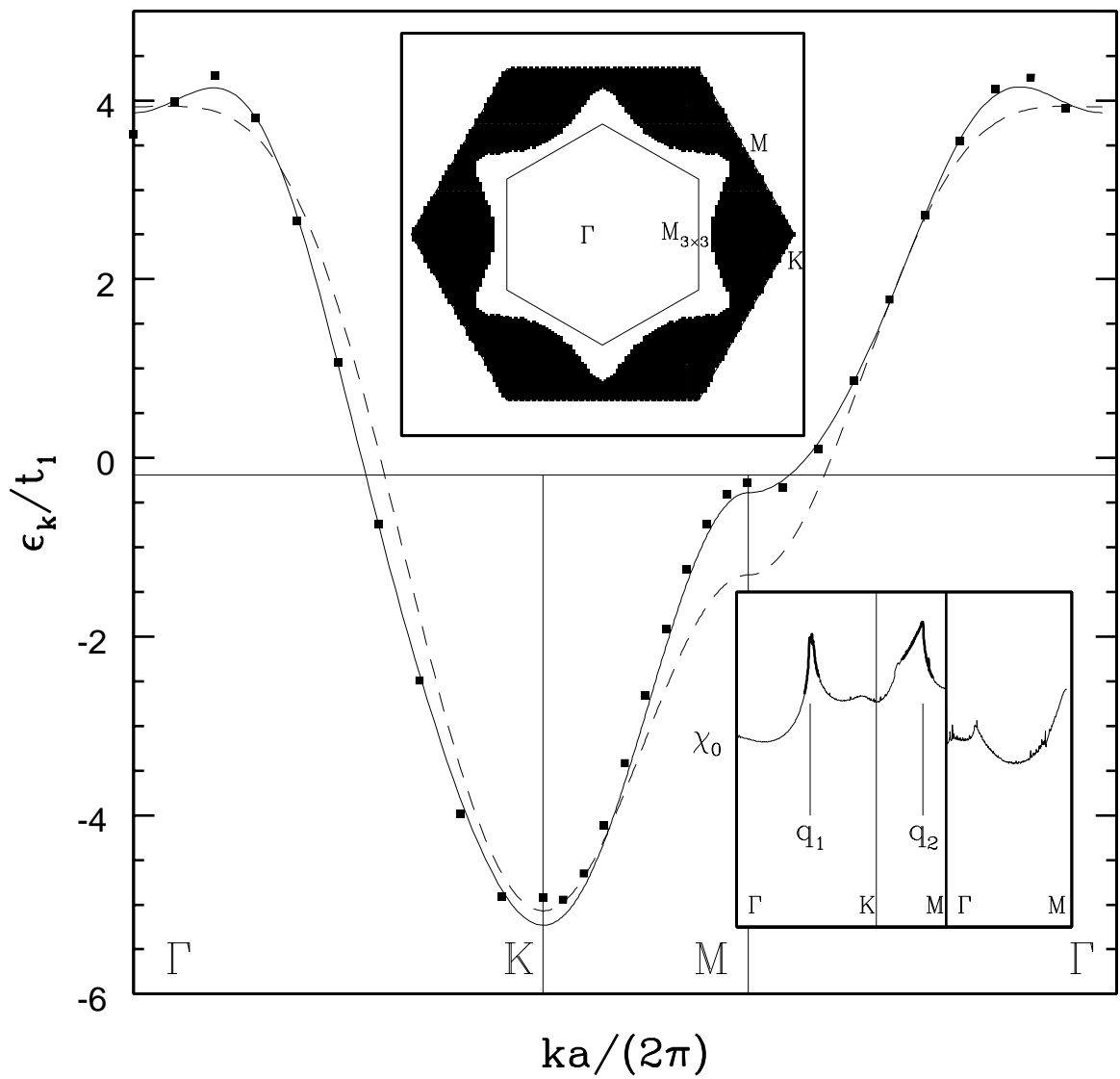
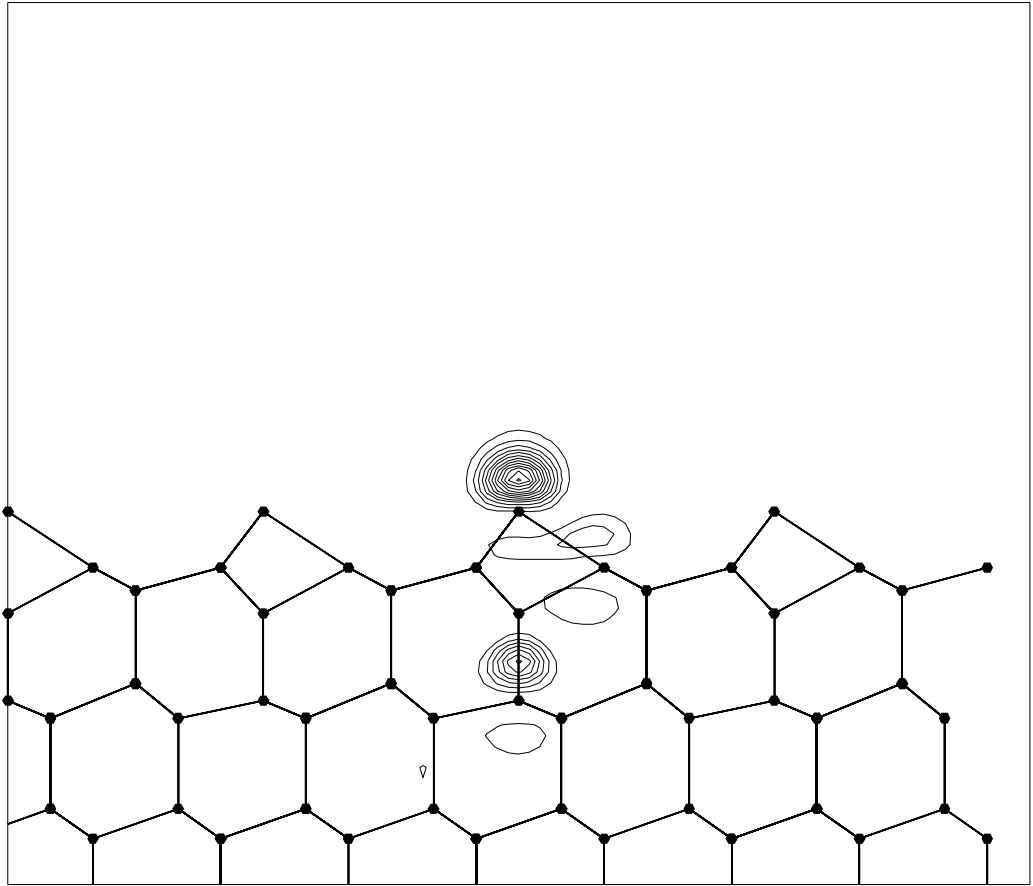
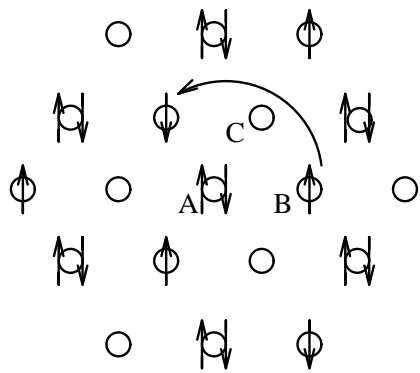
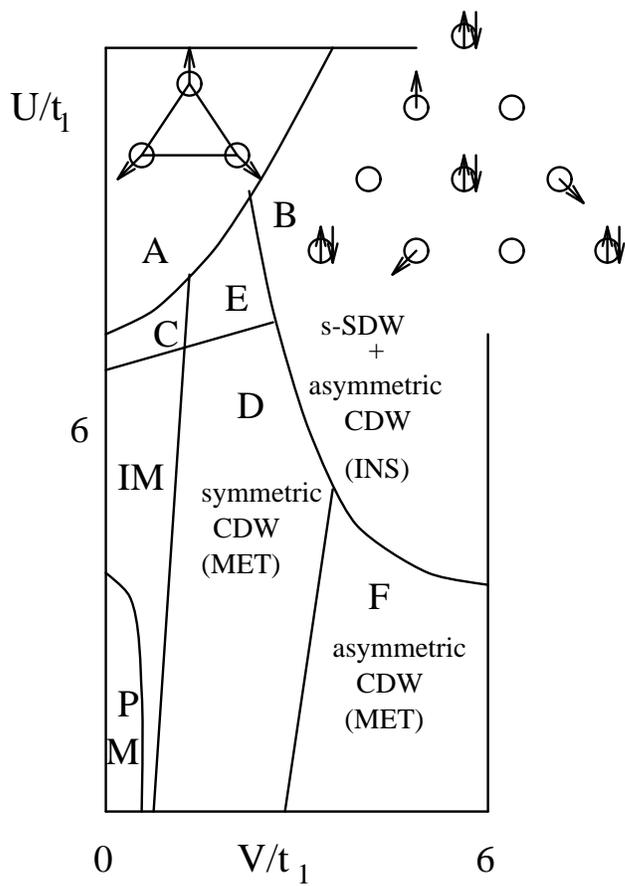
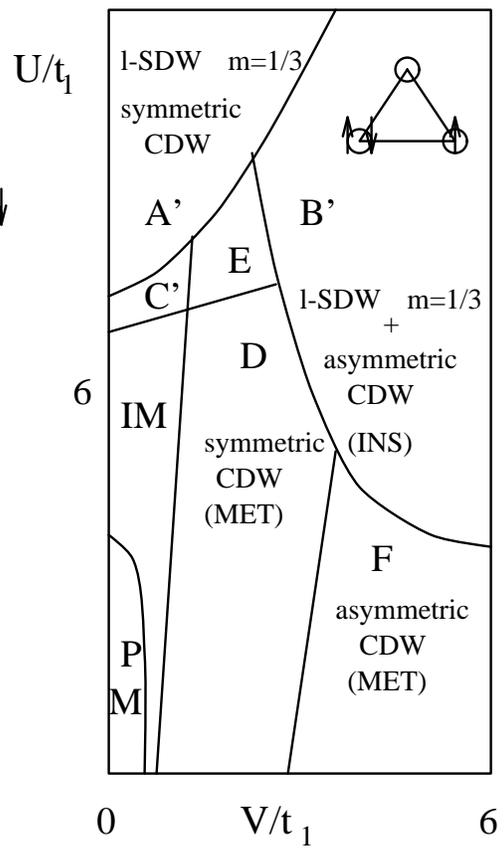
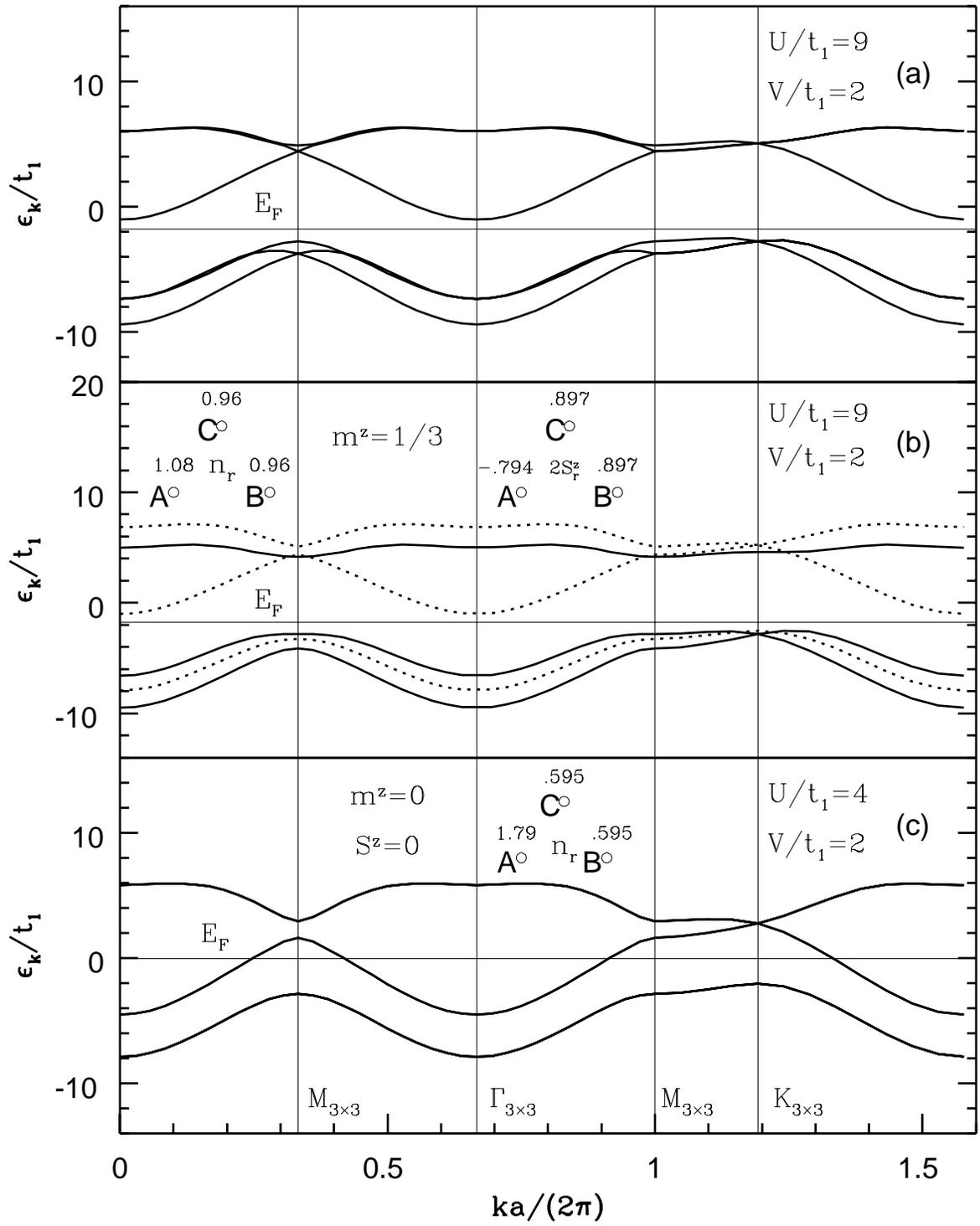
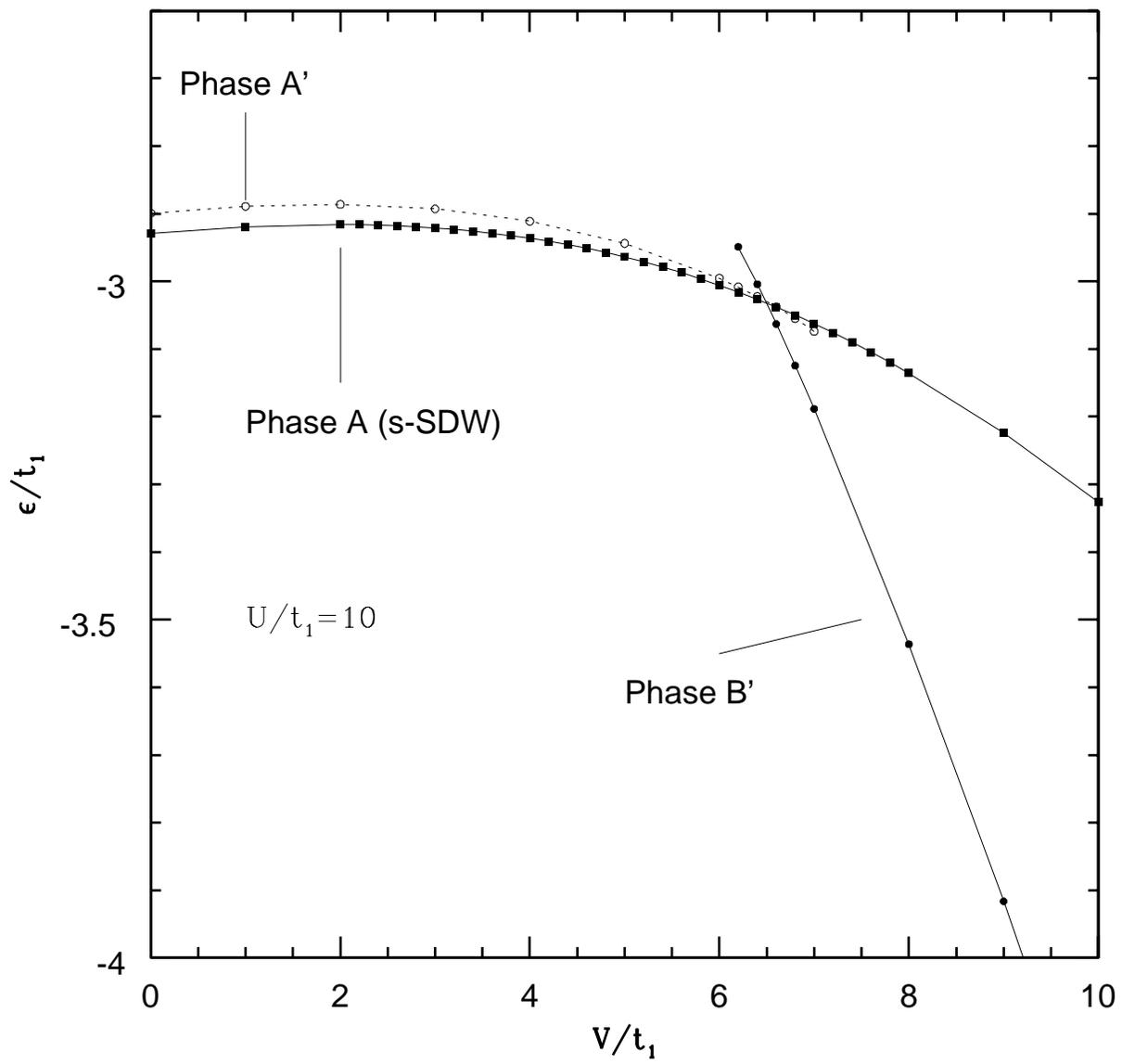
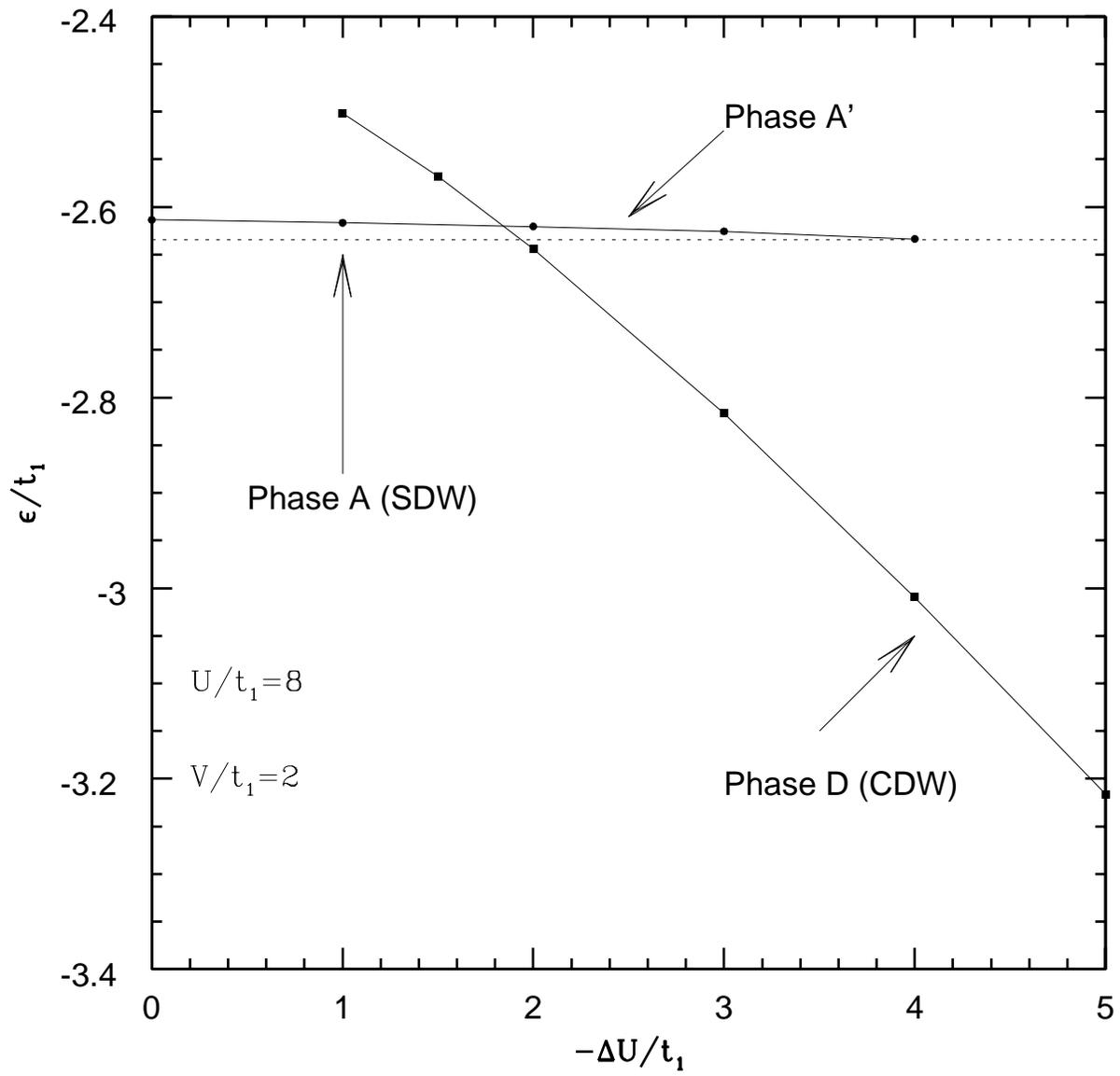
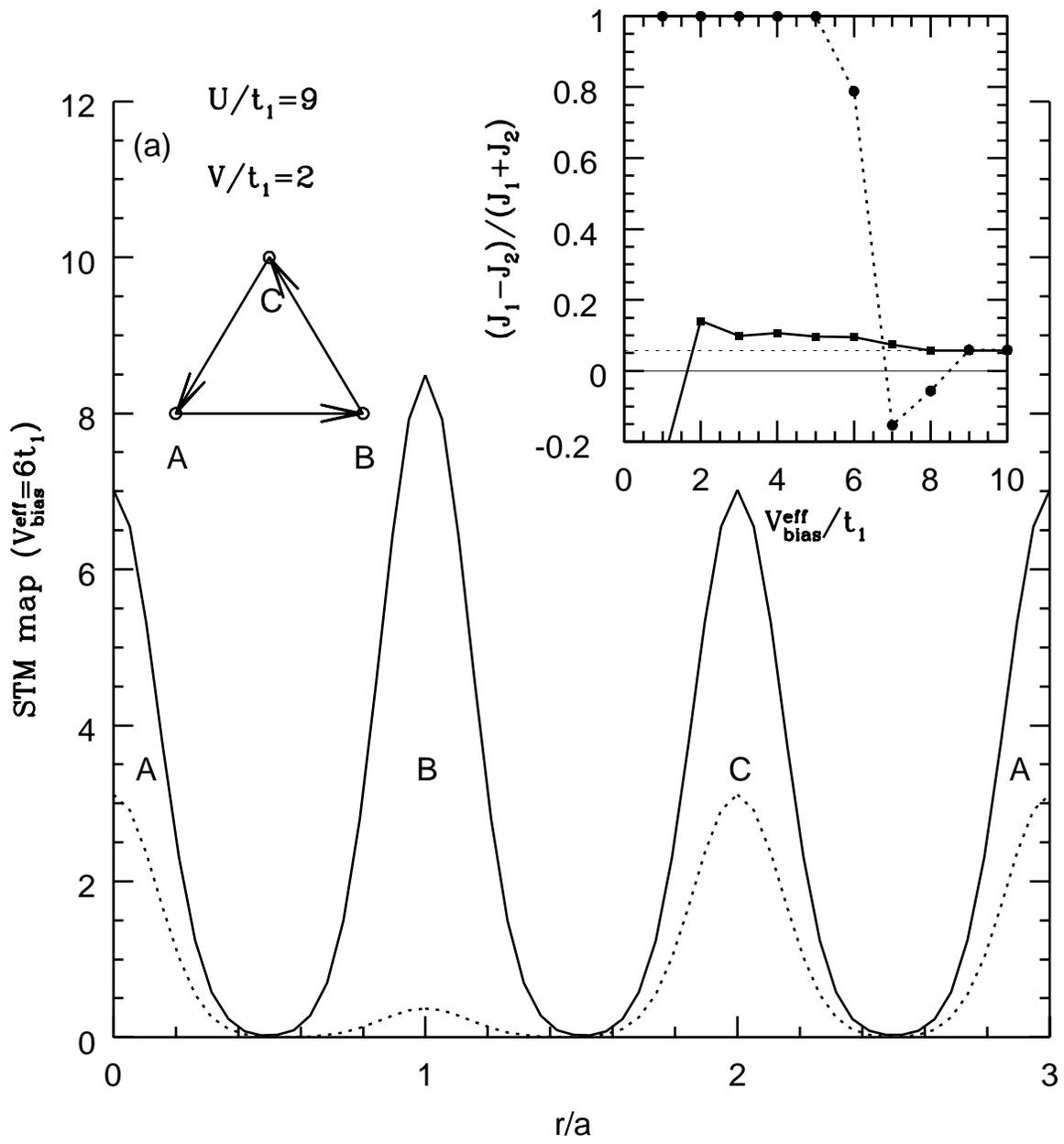
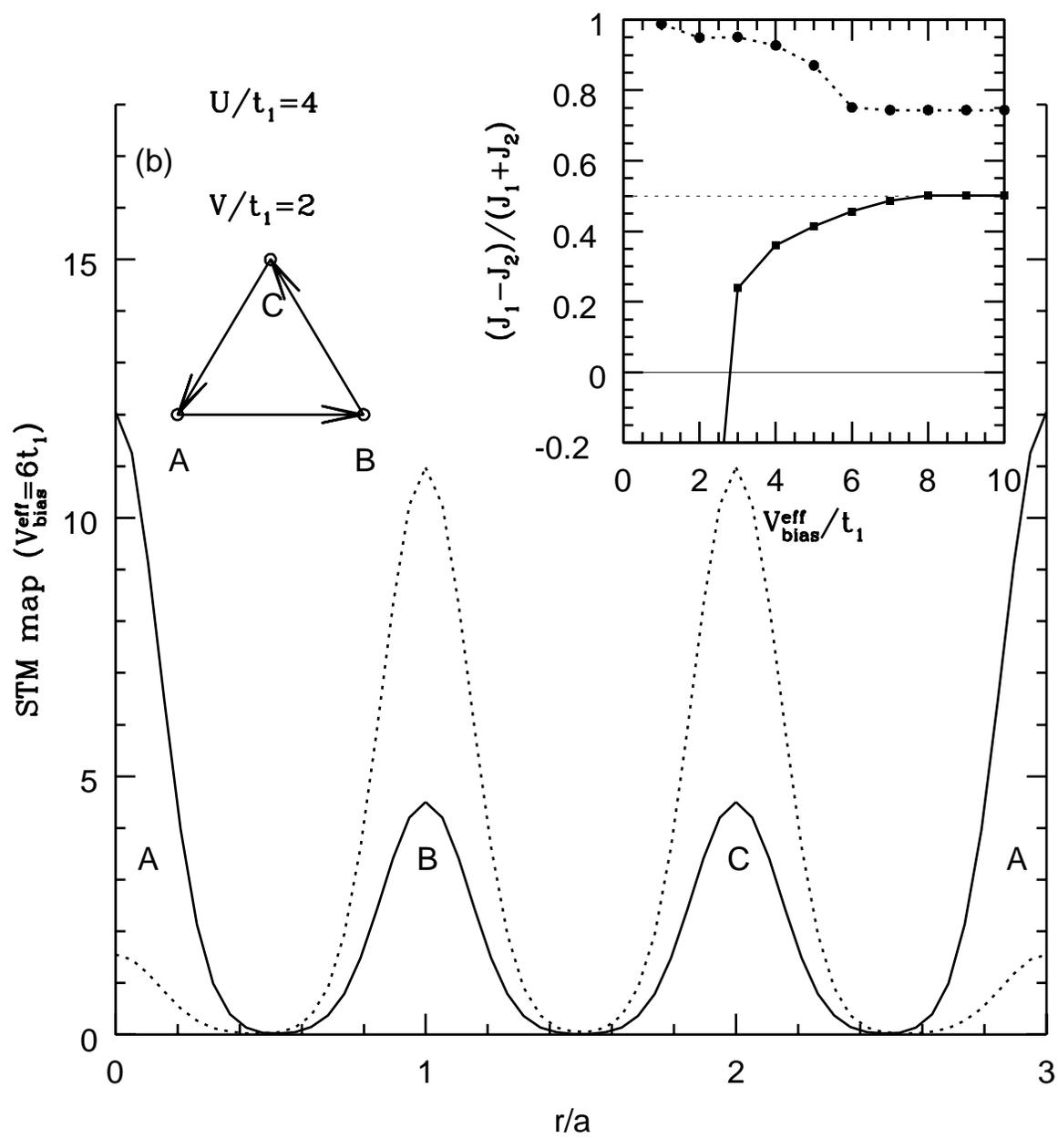
\end{document}